\documentclass[12pt]{article}
\usepackage{amsmath,amssymb}
\usepackage{graphicx}
\usepackage{color}
\begin{document}
\baselineskip=16pt
\begin{titlepage}
\begin{flushright}
{\small SU-HET-07-2014}
\end{flushright}
\vspace*{1.2cm}

\begin{center}

{\Large\bf Vanishing Higgs potential at the Planck scale in a singlet extension of the standard model}
\lineskip .75em
\vskip 1.5cm

\normalsize
{\large Naoyuki Haba},$^1$
{\large Hiroyuki Ishida},$^1$
{\large Kunio Kaneta},$^2$ and 
{\large Ryo Takahashi}$^1$

\vspace{1cm}
$^1${\it Graduate School of Science and Engineering, Shimane University, 

Matsue, Shimane 690-8504, Japan}\\
$^2${\it ICRR, University of Tokyo, Kashiwa, Chiba 277-8582, Japan}\\

\vspace*{10mm}

{\bf Abstract}\\[5mm]
{\parbox{13cm}{\hspace{5mm}
We discuss the realization of a vanishing effective Higgs potential at the Planck scale, which is required by the multiple-point criticality principle (MPCP), in the standard model with singlet scalar dark matter and a right-handed neutrino. We find the scalar dark matter and the right-handed neutrino play crucial roles for realization of the MPCP, where a neutrino Yukawa becomes effective above the Majorana mass of the right-handed neutrino. Once the top mass is fixed, the MPCP at the (reduced) Planck scale and the suitable dark matter relic abundance determine the dark matter mass, $m_S$, and the Majorana mass of the right-handed neutrino, $M_R$, as $8.5~(8.0)\times10^2~{\rm GeV}\leq m_S\leq1.4~(1.2)\times10^3~{\rm GeV}$ and $6.3~(5.5)\times10^{13}~{\rm GeV}\leq M_R\leq1.6~(1.2)\times10^{14}~{\rm GeV}$ within current experimental values of the Higgs and top masses. This scenario is consistent with current dark matter direct search experiments, and will be checked 
by future experiments such as LUX with further exposure and/or the XENON1T.
}}

\end{center}
\end{titlepage}

\section{Introduction}

The Higgs particle was discovered at the LHC 
experiment~\cite{Chatrchyan:2013lba,CMS}, but one finds no evidence to support 
the existence of physics beyond the standard model (SM) so far. 
Thus, the question ``How large is new physics scale?'' is important for the SM and new 
physics. One simple answer is that the SM is valid up to the Planck scale; i.e., 
there is no new physics between the electroweak (EW) and the Planck scales. In 
that case, the current experimental values of the Higgs and top masses might 
imply a vanishing effective Higgs potential at the Planck scale. In fact, 
there are intriguing researches about this possibility. For instance, 
Ref.~\cite{Froggatt:1995rt} proposed the multiple-point criticality principle 
(MPCP). This principle means that there are two degenerate vacua in the SM Higgs
 potential, $V(v)=V(M_{\rm pl})=0$ with $V'(v)=V'(M_{\rm pl})=0$, where $V$ is the effective 
Higgs potential, $v$ is the vacuum expectation value (VEV) of the Higgs, and 
$M_{\rm pl}$ is the Planck scale. One is at the EW scale where we live, and another
 is at the Planck scale, which can be realized by the Planck-scale boundary 
conditions (BCs) of the vanishing Higgs self-coupling [$\lambda(M_{\rm pl})=0$] 
and its $\beta$ function [$\beta_\lambda(M_{\rm pl})=0$]. As a result, 
Ref.~\cite{Froggatt:1995rt} pointed out that the principle predicts 
a $135\pm9~{\rm GeV}$ Higgs mass and a $173\pm5~{\rm GeV}$ top mass, which are close
 to the experimental values but not the current center values. Furthermore, an 
asymptotic safety scenario of gravity~\cite{Shaposhnikov:2009pv} predicted 
$126~{\rm GeV}$ Higgs mass with a few GeV uncertainty, and this scenario also 
pointed out $\lambda(M_{\rm pl})\simeq0$ and $\beta_\lambda(M_{\rm pl})\simeq0$ 
(see also Refs.~\cite{Holthausen:2011aa}-\cite{Spencer-Smith:2014woa} for more recent 
analyses). In this paper, we discuss the realization of a vanishing effective Higgs 
potential at the Planck scale, which  is required by the MPCP, in the SM with singlet scalar dark matter (DM) and a 
right-handed neutrino. 

An important motivation of the gauge singlet extension of the SM is to explain DM and the tiny active neutrino mass. In this extension, the 
scalar particle can be DM when it has odd parity under an additional $Z_2$ 
symmetry~\cite{Silveira:1985rk} (see 
also Refs.~\cite{McDonald:1993ex}-\cite{Boucenna:2014uma}). The right-handed Majorana 
neutrino can generate the tiny active neutrino mass via the type-I seesaw 
mechanism. Once the scalar (right-handed neutrino) is added to the SM, an 
additional positive (negative) contribution appears in $\beta_\lambda$.\footnote{See 
also Refs.~\cite{Davoudiasl:2004be,HKT,HKT2} for researches of the vacuum stability 
and the coupling perturbativity in the SM with scalar DM, 
and Refs.~\cite{Haba:2014zda,Hamada:2014xka,Ko:2014eia,Haba:2014zja} for explaining 
the recent BICEP2 result~\cite{Ade:2014xna} in the framework of the Higgs 
inflation~\cite{Bezrukov:2007ep} with gauge singlet fields.} In addition, since 
it is difficult to reproduce the 126 GeV Higgs mass and the $173.34\pm0.76$ GeV top pole
 mass~\cite{ATLAS:2014wva} at the same time under the MPCP at the Planck scale 
in the SM, it is intriguing to study whether the principle can be realized with 
the center values of the Higgs and top masses in the singlet extension of the 
SM, or not.

In this paper, we discuss the realization of the vanishing effective Higgs potential at the Planck scale, which is required by the MPCP, in the SM with singlet scalar DM and the right-handed neutrino. Intriguingly, both the scalar DM and the right-handed neutrino are necessary to realize the MPCP which predicts the DM mass $m_S$ and the Majorana mass of the right-handed neutrino $M_R$: $8.5~(8.0)\times10^2~{\rm GeV}\leq m_S\leq1.4~(1.2)\times10^3~{\rm GeV}$ and $6.3~(5.5)\times10^{13}~{\rm GeV}\leq M_R\leq1.6~(1.2)\times10^{14}~{\rm GeV}$ within current experimental values of the Higgs and top masses.


\section{Singlets extension of the SM}
The relevant Lagrangians of the singlet extension of the SM are given by
 \begin{align}
  &\mathcal{L}        
   =\mathcal{L}_{\rm SM}+\mathcal{L}_{\rm singlets}, \\
  &\mathcal{L}_{\rm SM} 
   \supset-\lambda\left(|H|^2-\frac{v^2}{2}\right)^2, \label{S} \\
  &\mathcal{L}_{\rm singlets}      
   =-\frac{\bar{m}_S^2}{2}S^2-\frac{k}{2}|H|^2S^2
             -\frac{\lambda_S}{4!}S^4-\left(\frac{M_R}{2}\overline{N^c}N
             +y_N\overline{L}\tilde{H}N+c.c.\right)+({\rm kinetic~term}),
 \end{align}
where the SM Lagrangian including the effective 
Higgs potential is given by $\mathcal{L}_{\rm SM}$, and $H$ is the Higgs doublet ($\tilde{H}\equiv-i\sigma_2H^\ast$), 
$S$ is a gauge singlet real scalar field, $L$ is the left-handed lepton doublet 
of the SM, $N$ is the right-handed neutrino, $y_N$ is the neutrino Yukawa 
coupling, and $M_R$ is the Majorana mass of the right-handed neutrino. In the 
model, since only the singlet real scalar is assumed to have odd parity under an
 additional $Z_2$ symmetry, it can be DM with suitable mass and couplings. The 
DM mass is given by $m_S=\sqrt{\bar{m}_S^2+kv^2/2}$. The right-handed 
neutrino generates the small active neutrino mass through the type-I seesaw 
mechanism.

We utilize the renormalization group equations (RGEs) at two-loop level in this 
model, which were first given in Ref.~\cite{Haba:2014zja}. Here, we mention the 
features of RGE runnings of the scalar quartic couplings at the two-loop level:
\begin{enumerate}
\item Since the $\beta$ function of $k$ is proportional to $k$ itself, an evolution of $k$ is tiny when $k(M_Z)$ is close to zero. Note that $k(M_Z)\rightarrow0$ is the SM limit.

\item $\lambda(\mu)$ becomes negative within $\mathcal{O}(10^{10})~{\rm GeV}\lesssim\mu\leq M_{\rm pl}$ when the experimental center values of the Higgs and top masses are taken; this is known as the vacuum instability or meta-stability in the SM. This is induced from the dominant negative contribution of the top Yukawa coupling, $-6y^4$. NNLO computations~\cite{Degrassi:2012ry} indicate that $\lambda(\mu)$ can be positive within $M_Z\leq\mu\leq M_{\rm pl}$ for the Higgs mass as $127~{\rm GeV}\leq m_h\leq130~{\rm GeV}$ with a top mass of $M_t=173.1\pm0.6~{\rm GeV}$ or $171.3~{\rm GeV}\leq M_t\leq171.7~{\rm GeV}$, with $m_h=126~{\rm GeV}$ (see also Ref.~\cite{Spencer-Smith:2014woa}).

\item The RGE evolution of $\lambda$ can be raised by the additional positive term $+k^2/2$ in the $\beta$ function of $\lambda$. There is also a negative contribution $-2y_N^4$ to the $\beta$ function of $\lambda$ from the neutrino Yukawa coupling, which pushes down the RGE evolution of $\lambda$. We will investigate whether the MPCP can be realized by considering these two contributions in the model, or not. 
\end{enumerate}

\section{Multiple point criticality principle in singlets extension of the SM}

The MPCP requires that there exist two degenerate vacua in the effective Higgs potential. 
One is at the EW scale where we live and another is at the Planck scale. This  
principle is described as $V(v)=V(M_{\rm pl})=0$. In terms of $\lambda$ and 
$\beta_\lambda$, this principle is written as
 \begin{eqnarray}
  \lambda(M_{\rm pl})=0,~~~\beta_\lambda(M_{\rm pl})=0, \label{M}
 \end{eqnarray}
which is obtained from the stationary condition, $V'(H)=0$. The conditions cannot be realized in the SM within the current experimental ranges of top and Higgs masses; i.e., the MPCP in the SM requires a lighter top mass and/or a heavier Higgs mass~\cite{Froggatt:1995rt}. Thus, we need to consider an extension of the SM anyhow.
 
We investigate a realization of Eq.~(\ref{M}) in the singlet extension of the SM by 
solving the two-loop-level RGEs. 
The scalar DM (neutrino Yukawa) coupling lifts up (pushes down)
 the running of $\lambda$. Thanks to these two contributions in this extension, the positive contribution from the scalar to $\beta_\lambda$ can avoid the metastable vacuum of the SM with the current experimental values of the top and the Higgs masses. The scalar contribution becomes dominant in $\beta_\lambda$ at the Planck scale, which can realize $\lambda(M_{\rm pl})>0$ and $\beta_\lambda(M_{\rm pl})>0$. And a negative contribution from the neutrino Yukawa coupling to $\beta_\lambda$ above the Majorana mass scale can successfully achieve $\lambda(M_{\rm pl})=0$ and $\beta_\lambda(M_{\rm pl})=0$. This is the essence of the realization of the MPCP in this singlet extension of the SM, and the realization is nontrivial. 
For the RGEs, decoupling effects of the scalar and the right-handed neutrino  should be taken into account below their mass scales by taking away their relevant couplings from the corresponding $\beta$ functions. In particular, it is very important that the neutrino Yukawa becomes effective above the Majorana mass of the right-handed neutrino. For the neutrino sector, the active neutrino mass is induced from the seesaw mechanism [$m_\nu=y_N^2v^2/(2M_R)$] and is taken as $m_\nu=0.1$ eV. With these relations, the value of $y_N(M_Z)$ is given by $y_N(M_Z)=\sqrt{2m_\nu M_R}/v$. This is an example in which one active neutrino mass is obtained. Two other neutrino masses can also be effective in the RGE analyses, but here we assume that other neutrino Yukawa couplings are small enough to be neglected in the analyses.

\begin{figure}
\begin{center}
\begin{tabular}{cc}
(a) & (b) \\
\includegraphics[scale=0.79]{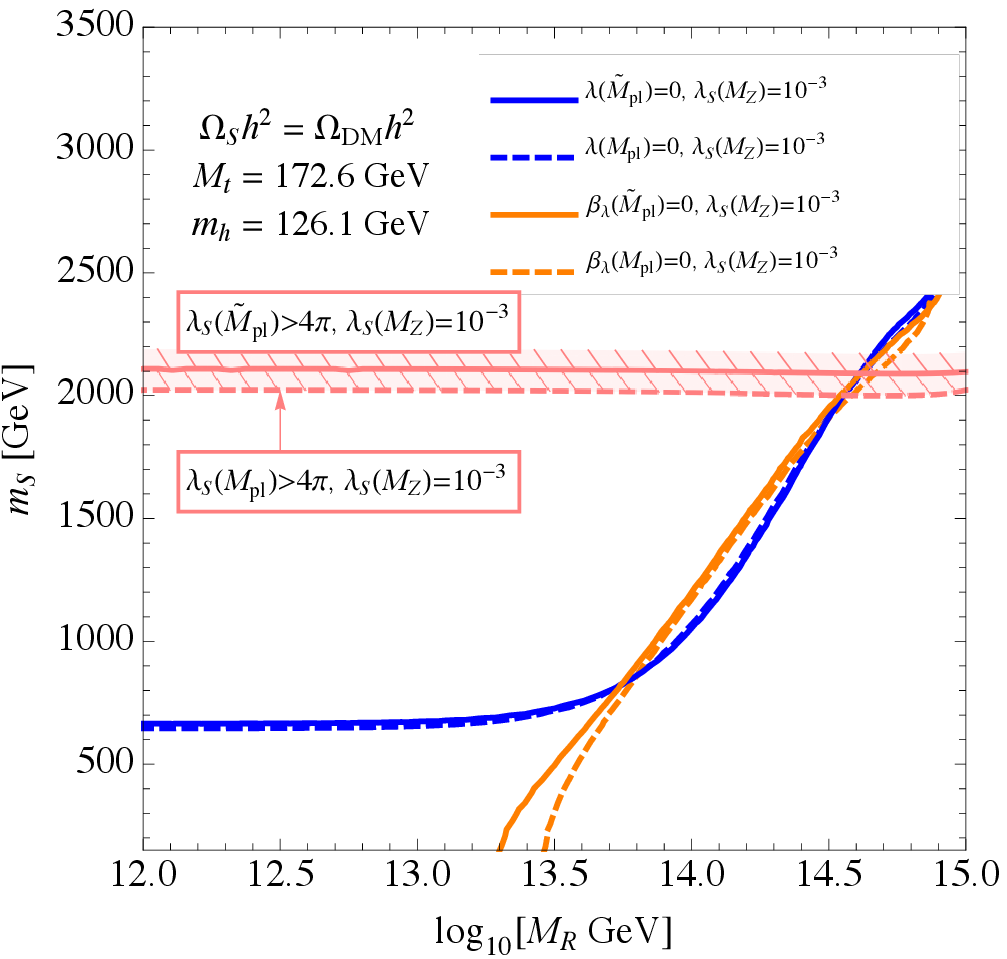} &
\includegraphics[scale=0.79]{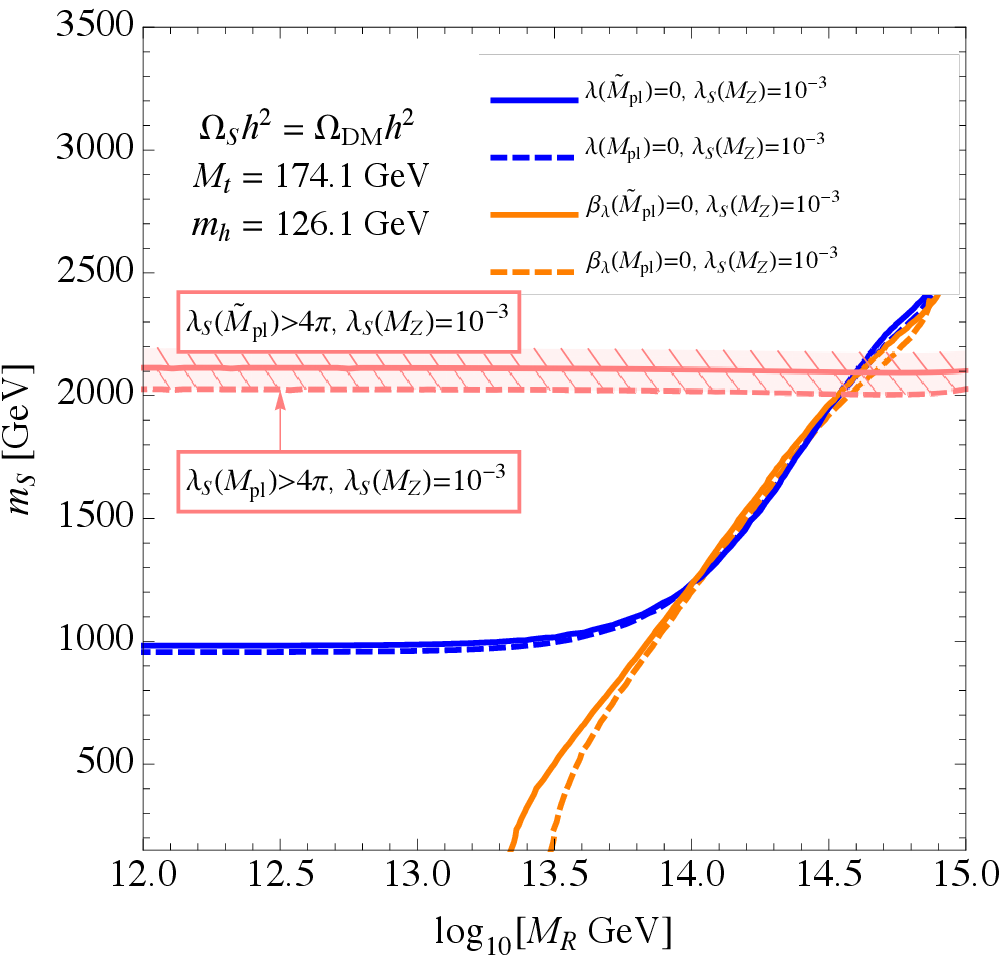} \\
(c) & (d) \\
\includegraphics[scale=0.79]{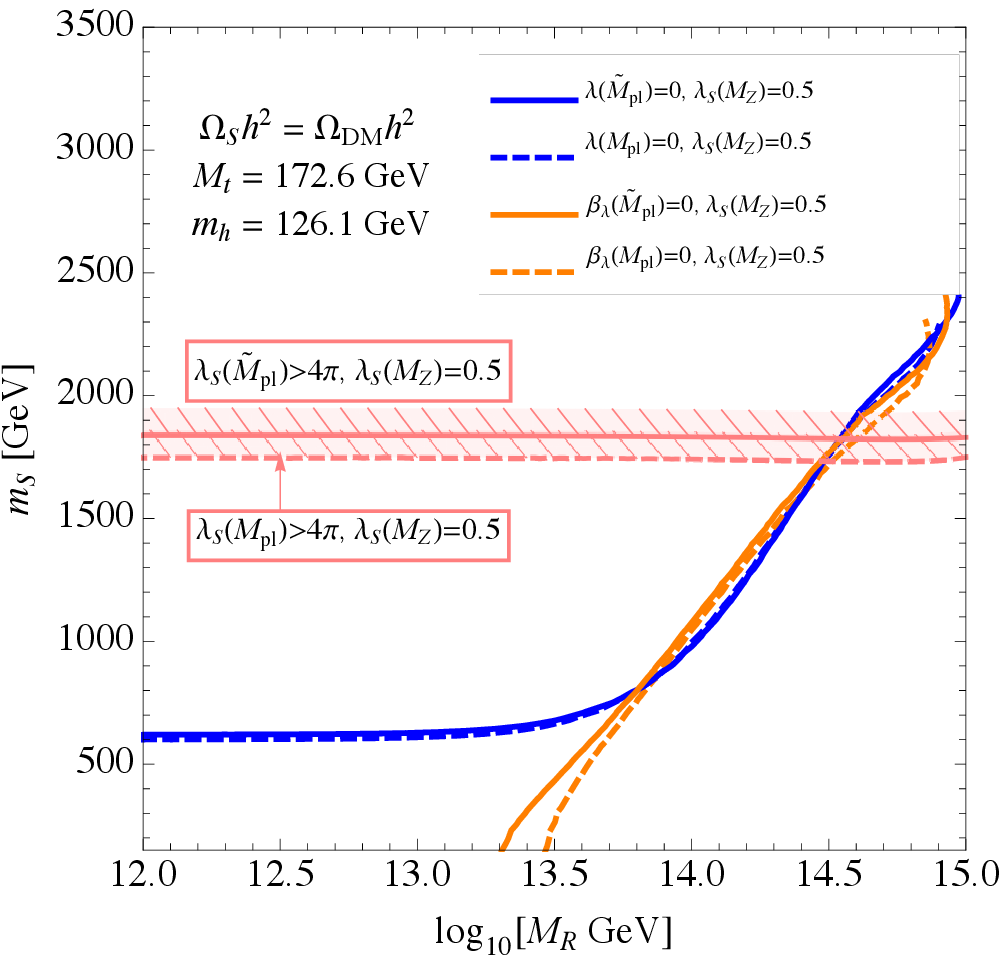} &
\includegraphics[scale=0.79]{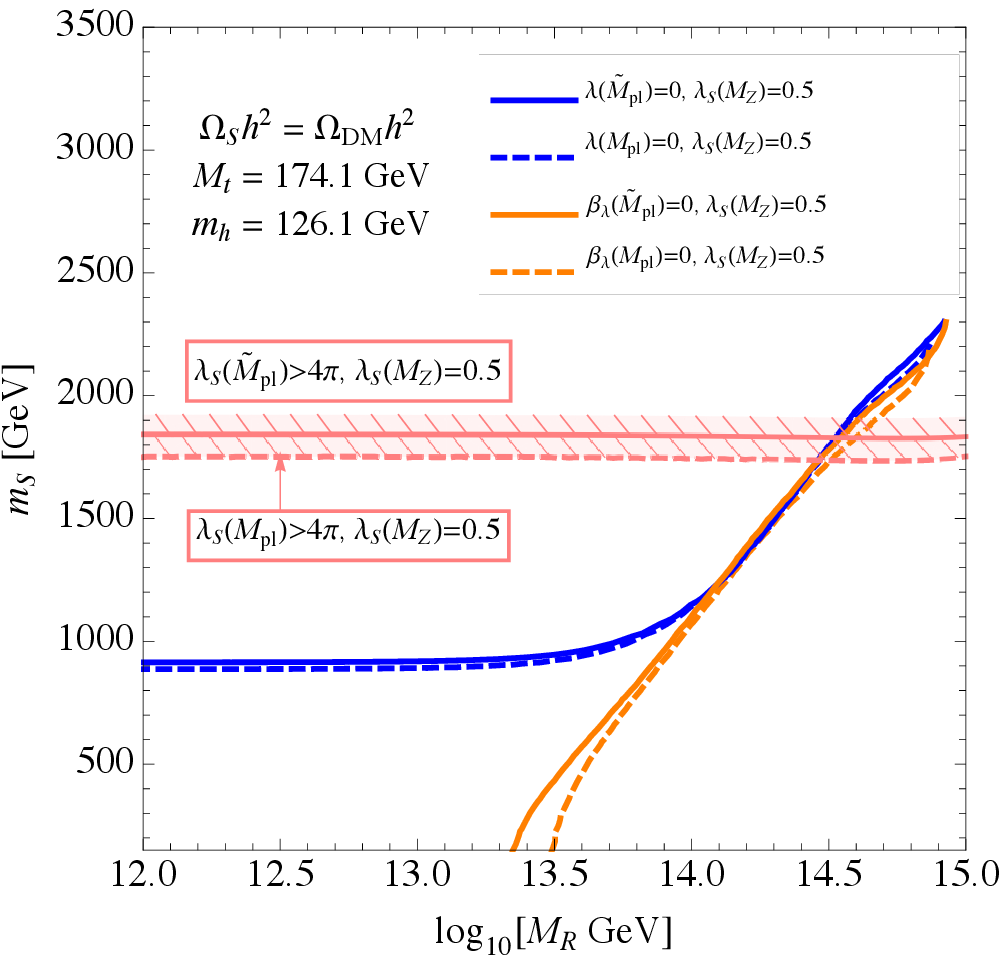} 
\end{tabular}
\end{center}
\caption{Numerical results for the realization of the MPCP [$\lambda=0$ (blue curve) and $\beta_\lambda=0$ (orange curve)]. The contours for the two conditions at the 
Planck $M_{\rm pl}$ and the reduced Planck $\tilde{M}_{\rm pl}$ scales are shown by 
the dashed and solid curves in all figures, respectively. Figures (a), (b), (c), and (d) correspond to the cases of $(M_t~[{\rm GeV}],\lambda_S(M_Z))=(172.6,10^{-3})$, 
$(174.1,10^{-3})$, $(172.6,0.5)$, and $(174.1,0.5)$, respectively.
}
\label{fig1}
\end{figure}

Our results are summarized in 
Fig.~\ref{fig1},  
where the conditions of $\lambda=0$ and $\beta_\lambda=0$ are depicted 
by blue and orange curves, respectively.
We analyze the realization of the MPCP at both the Planck $M_{\rm pl}$ and the reduced Planck scales $\tilde{M}_{\rm pl}$, which are shown by the dashed and solid curves in all figures, respectively. In the regions above (below) the blue and orange solid curves, 
$\lambda(M_{\rm pl})>0$ and $\beta_\lambda(M_{\rm pl})>0$ [$\lambda(M_{\rm pl})<0$ and $\beta_\lambda(M_{\rm pl})<0$], 
respectively. These correspondences are the same for the case of the reduced 
Planck scale. Figures~\ref{fig1} (a)-\ref{fig1} (d) are the cases of $(M_t~[{\rm GeV}],\lambda_S(M_Z))=(172.6,10^{-3})$, $(174.1,10^{-3})$, $(172.6,0.5)$, and $(174.1,0.5)$, respectively, with $\Omega_Sh^2=\Omega_{\rm DM}h^2=0.12$~\cite{Ade:2013zuv}, where $\Omega_S$ is the density parameter of $S$, $\Omega_{\rm DM}$ is for DM, and $h$ is the Hubble constant. Since $S$ is DM in the model,  
the value of $k(M_Z)$ is determined by $m_S$ and $\Omega_S$. We utilize {\tt micrOMEGAs} \cite{Belanger:2013oya} to estimate the relic 
abundance of $S$, and we take the Higgs mass as 126.1 GeV in the 
calculation.\footnote{We also take the strong coupling as $\alpha_s=0.1184$. For
 the matching terms of $y_t$ and $\lambda$ at the top pole mass scale, we take 
two-loop results, shown in e.g. Refs.~\cite{Holthausen:2011aa,Degrassi:2012ry}.} In the region above the pink dashed and
 solid lines, self-coupling $\lambda_S$ exceeds $4\pi$ at the Planck and reduced
 Planck scales, respectively, while perturbative calculation is 
valid in the parameter space below the pink lines. At an intersection point of 
the blue and orange solid (dashed) curves below the horizontal pink solid 
(dashed) line, the MPCP can be satisfied within the experimentally allowed region of
 the Higgs, top, and DM masses with suitable scalar quartic couplings up to the 
(reduced) Planck scale. One can really see that there are some intersection 
points in Fig.~\ref{fig1}. We mention parameter dependences for the realization 
of the MPCP as follows:
\begin{enumerate} 
\item When $M_R$ is relatively light, as $M_R<\mathcal{O}(10^{13})$~GeV, the contribution from the neutrino Yukawa coupling to $\beta_\lambda$ is negligible. Thus, once $M_t$ is 
fixed, $\lambda=0$ is realized by taking suitable value for only $m_S$. This is shown by flat regions of blue curves in the figures. In this region, $\beta_\lambda(M_{\rm pl})$ is always positive. A similar case, i.e. the decoupling limit of the right-handed neutrino $y_N\rightarrow0$, was discussed in Ref.~\cite{HKT2}, and our analysis is consistent with the results of Ref.~\cite{HKT2}. 
 
\item When $M_R$ becomes large, we can successfully achieve $\beta_\lambda(M_{\rm pl})=0$ with $\lambda(M_{\rm pl})=0$. The correlation between $m_S$ and $M_R$ is seen in the slanting regions of the blue curves. One can see that a larger value of $M_R$ is required to balance with the large scalar contribution.

\item Regarding the coupling perturbativity of $\lambda_S(M_{\rm pl})$, it strongly depends on values of $k(M_Z)$ and $\lambda_S(M_Z)$ but not on $M_R$, because the neutrino Yukawa does not contribute to $\beta_{\lambda_S}$ at the one-loop level. Thus, when one takes a larger $\lambda_S(M_Z)$, the bound of the coupling perturbativity of $\lambda_S(M_{\rm pl})$ becomes severe for $m_S$.

\item For heavier $M_t$, the MPCP can be realized in heavier $m_S$, or equivalent to larger $k(M_Z)$ [compare Figs.~\ref{fig1}(a) and \ref{fig1}(b), or Figs.~\ref{fig1}(c) and \ref{fig1}(d)]. This is because the dominant negative contribution from the top Yukawa coupling in $\beta_\lambda$ should be canceled by the positive contribution from $k$.

\item For larger $\lambda_S(M_Z)$, the MPCP is satisfied in lighter $m_S$ 
[compare Fig.~\ref{fig1}(a) and \ref{fig1}(c), or Figs.~\ref{fig1}(b) and \ref{fig1}(d)]. Since $\lambda_S$ coupling gives a positive contribution to the $\beta$ function of $k$, $k$ grows more rapidly for larger $\lambda_S(M_Z)$. As a result, smaller $k(M_Z)$ (or $m_S$) is favored for canceling the negative contribution from the top Yukawa coupling in a larger $\lambda_S(M_Z)$ case.
\end{enumerate}

\begin{figure}
\begin{center}
\includegraphics[scale=0.3]{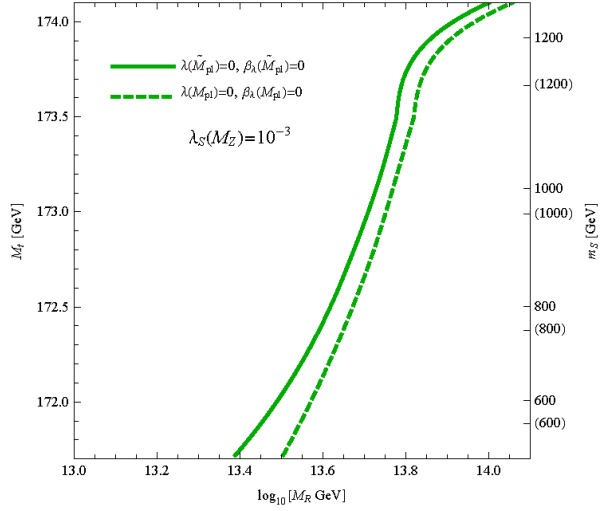}
\end{center}
\caption{The positions of the intersection points in the [$M_R,M_t(\mbox{or }m_S)$] plane for the $\lambda_S(M_Z)=10^{-3}$ case. The solid and dashed curves indicate the MPCP solutions at $\tilde{M}_{\rm pl}$ and $M_{\rm pl}$, respectively. The values in the parentheses 
on the $m_S$ axis correspond to values in the case of the MPCP at the Planck
 scale while the $m_S$ values without parentheses are for the case of MPCP at 
the reduced Planck scale.
}
\label{fig1-3}
\end{figure}
Next, Fig.~\ref{fig1-3} shows the positions of the intersection points 
in the [$M_R,M_t(\mbox{or }m_S)$] plane for the $\lambda_S(M_Z)=10^{-3}$ case.\footnote{One might 
also find another intersection point around $m_S\sim2.0\times10^3$ GeV in each figure [($m_S~[{\rm GeV}],M_R~[{\rm GeV}])=(1.9\times10^3,3.0\times10^{14})$, $(1.8\times10^3,2.6\times10^{14})$, $(1.6\times10^3,2.5\times10^{14})$, and $(1.4\times10^3,1.7\times10^{14})$ in Figs.~\ref{fig1}(a), \ref{fig1}(b), \ref{fig1}(c), and \ref{fig1}(d), respectively]. Since these points are close to the lines of the coupling perturbativity bound on $\lambda_S$, we do not consider these solutions around $m_S\sim2.0\times10^3$ GeV in this paper anymore. But this could also be the solution for the MPCP.} The solid and dashed curves indicate the MPCP solutions at $\tilde{M}_{\rm
 pl}$ and $M_{\rm pl}$, respectively. We can show that $m_S$ and $M_R$ have one-to-one 
correspondence ($m_S$ and $M_t$ also have one to one correspondence). When one 
takes larger $M_t$, larger $m_S$ and $M_R$ are required to achieve 
$\lambda=0$ and $\beta_\lambda=0$ at the same time. To summarize, there are seven independent parameters; i.e., five coupling constants ($\lambda$, $k$, $\lambda_S$, $y$, and $y_N$) and two mass scales of the singlets ($m_S$ and $M_R$), in the scalar and Yukawa sectors of the model, in which $\lambda$ is determined by $m_h$. The suitable DM relic abundance relates $k$ with $m_S$ and the seesaw mechanism relates $y_N$ with $M_R$. Thus, there are four independent parameters ($\lambda_S$, $y$, $m_S$, and $M_R$). When the top mass and $\lambda_S$ are fixed, the two conditions of the MPCP ($\lambda=0$ and $\beta_\lambda=0$) uniquely determine $m_S$ and $M_R$.\footnote{By extending the model, $m_S$ and $M_R$ could be induced dynamically from a dimensional transmutation, which could have a conformal or shift symmetry in the framework of conformal gravity as a UV theory.}

As our result, we find that the MPCP at the (reduced) Planck scale predicts the following mass
 regions:
\begin{eqnarray}
&&8.5~(8.0)\times10^2~{\rm GeV}\leq m_S\leq1.4~(1.2)\times10^3~{\rm GeV}, \label{ms1} \\
&&6.3~(5.5)\times10^{13}~{\rm GeV}\leq M_R\leq1.6~(1.2)\times10^{14}~{\rm GeV},
\end{eqnarray}
within $M_t=(172.6\mathchar`-174.1)$ GeV and $10^{-3}\leq\lambda_S(M_Z)\leq0.5$. They are obtained by maximal and minimal values of $m_S$ and $M_R$ on the intersection points of the two contours of the $\lambda=0$ and $\beta_\lambda=0$ lines which are located at ($m_S~[{\rm GeV}],M_R~[{\rm GeV}])=(8.6~(8.2)\times10^2,6.3~(5.5)\times10^{13})$, $(1.3~(1.2)\times10^3,1.1~(1.0)\times10^{14})$, $(8.5~(8.0)\times10^2,7.4~(6.3)\times10^{13})$, and $(1.4~(1.2)\times10^3,1.6~(1.2)\times10^{14})$ for the case of the MPCP at the (reduced) Planck scale shown in Figs.~\ref{fig1}(a), \ref{fig1}(b), \ref{fig1}(c), and \ref{fig1}(d), respectively.\footnote{There are also intersection points around $m_S\sim2.0\times10^3$ GeV in the 
reduced Planck case as ($m_S~[{\rm GeV}],M_R~[{\rm 
GeV}])=(2.1\times10^3,3.9\times10^{14})$, $(2.0\times10^3,3.5\times10^{14})$, 
$(1.8\times10^3,3.3\times10^{14})$, and $(1.7\times10^3,2.8\times10^{14})$ in 
Figs.~\ref{fig1}(a), \ref{fig1}(b), \ref{fig1}(c), and \ref{fig1}(d), respectively.}

\begin{figure}
\begin{center}
\includegraphics[scale=0.77]{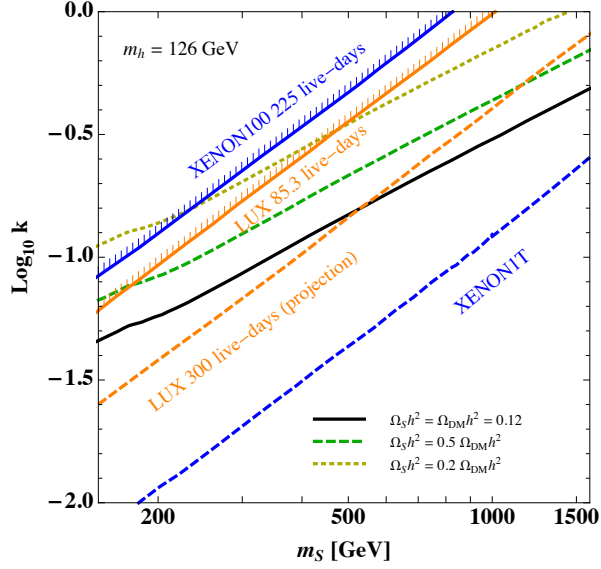}
\end{center}
\caption{The current 
experimental bounds on $m_S$ and $k$ (XENON100 225 live-days (blue solid line) and LUX 85.3 live-days (orange solid line)), and the 
future detectability by the LUX 300 day projection (orange dashed line) and the 
XENON1T (blue dashed line)~\cite{Akerib:2013tjd,Cline:2013gha}. The black solid curve indicates the contour of $\Omega_Sh^2=\Omega_{\rm DM}h^2=0.12$. (The green and yellow dashed curves correspond to $\Omega_Sh^2=0.5\Omega_{\rm DM}h^2$ and $0.2\Omega_{\rm DM}h^2$, respectively.)
}
\label{fig2}
\end{figure}

Finally, we draw Fig.~\ref{fig2}, which shows the current experimental bounds on $m_S$ and $k$ [XENON100 225, live-days (blue solid line) and LUX, 85.3 live-days (orange solid line)] and the future detectability by the LUX (orange dashed line) and the 
XENON1T (blue dashed line) experiments~\cite{Akerib:2013tjd,Cline:2013gha}. The black solid curve indicates the contour of $\Omega_Sh^2=\Omega_{\rm DM}h^2=0.12$. One can find that the DM mass region in Eq.~(\ref{ms1}) for the realization of the 
MPCP can be consistent with the current DM direct detection experiments, and it will be checked by future DM direct searches, e.g., the future XENON1T experiment.

\section{Summary and discussions}

We have discussed the realization of the vanishing effective Higgs potential at the Planck 
scale, which is required by the MPCP, in the SM with the singlet scalar DM and the right-handed neutrino. We have found that the scalar DM and the right-handed neutrino play crucial roles for realization of the MPCP, where the neutrino Yukawa becomes effective above the Majorana mass of the right-handed neutrino. Once the top mass is fixed, the MPCP at the (reduced) Planck scale and the suitable DM relic abundance determine the DM mass and Majorana mass of the right-handed neutrino as $8.5~(8.0)\times10^2~{\rm GeV}\leq m_S\leq1.4~(1.2)\times10^3~{\rm GeV}$ and $6.3~(5.5)\times10^{13}~{\rm GeV}\leq M_R\leq1.6~(1.2)\times10^{14}~{\rm GeV}$ within the current experimental values of the Higgs and top masses. The $m_S$ region is allowed by the current experimental results of the DM direct searches. Moreover, it is of importance that this scenario is testable by the future direct search experiments such as the LUX with further exposure and/or the XENON1T.

\begin{figure}
\begin{center}
\begin{tabular}{cc}
(a) & (b) \\
\includegraphics[scale=0.45]{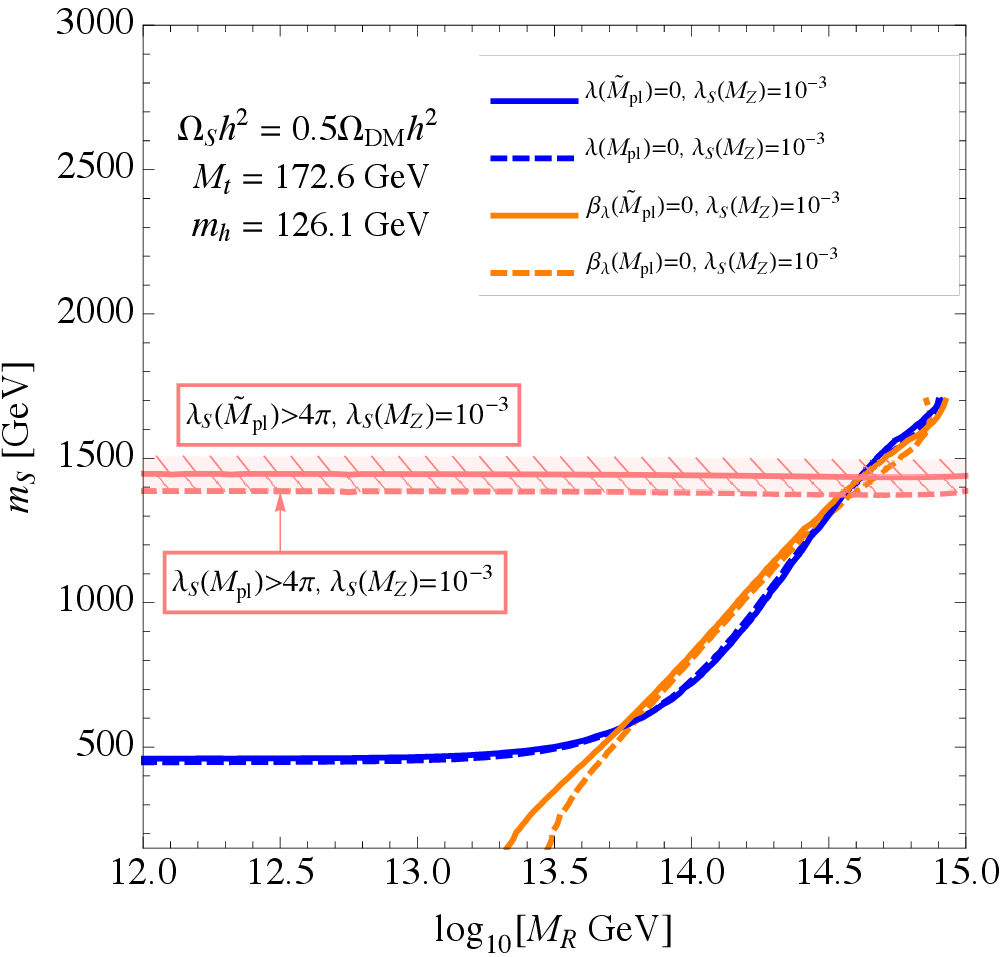} &
\includegraphics[scale=0.45]{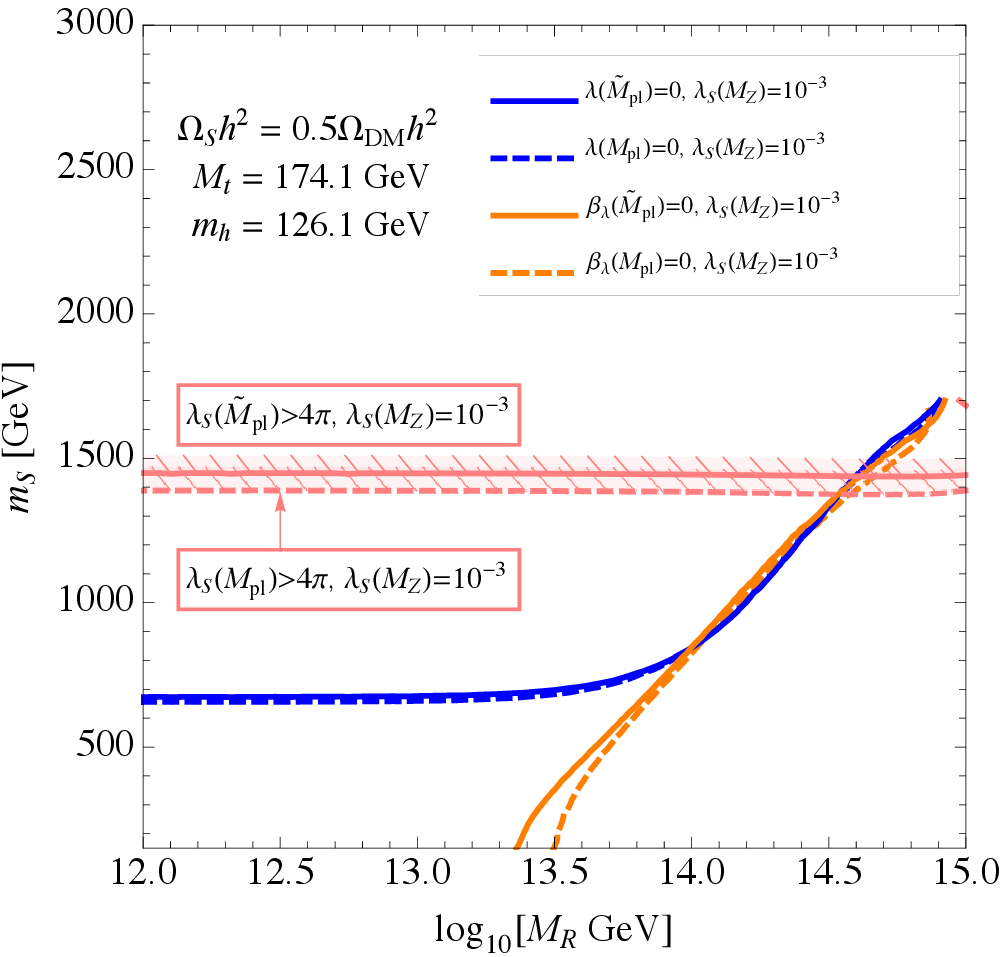} \\
(c) & (d) \\
\includegraphics[scale=0.45]{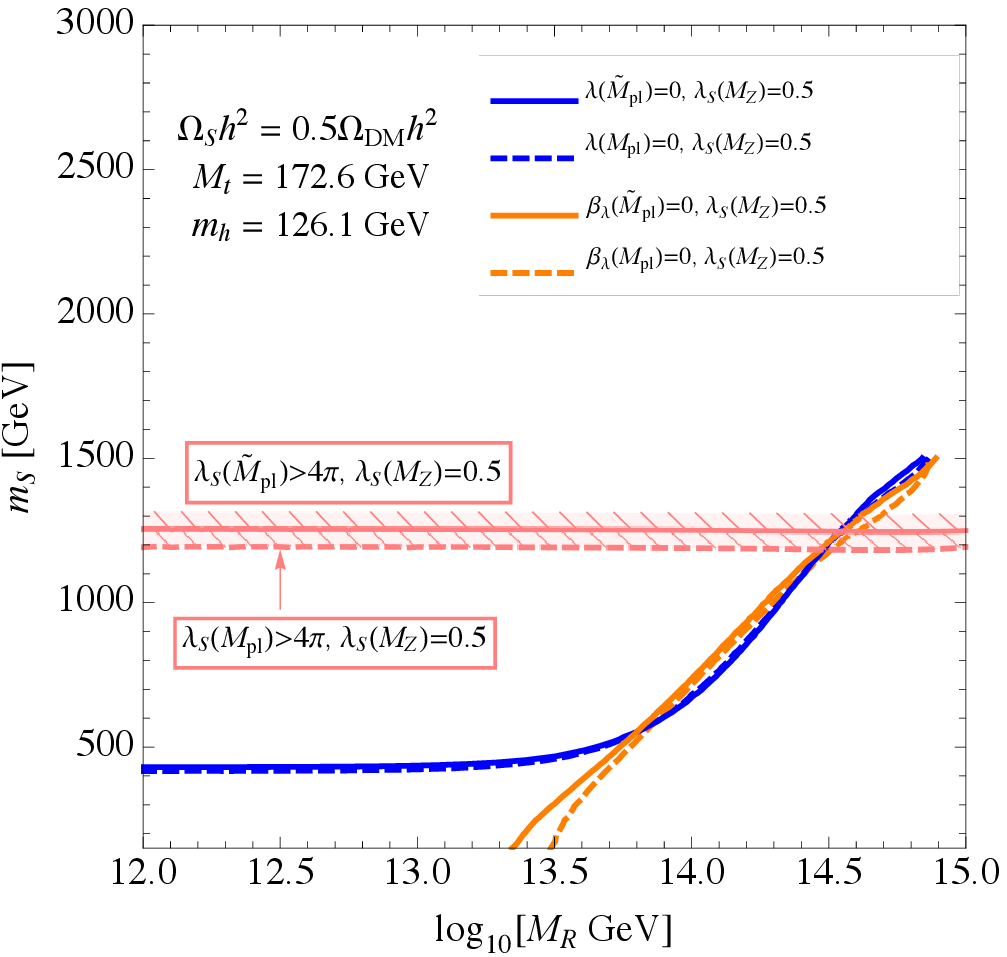} &
\includegraphics[scale=0.45]{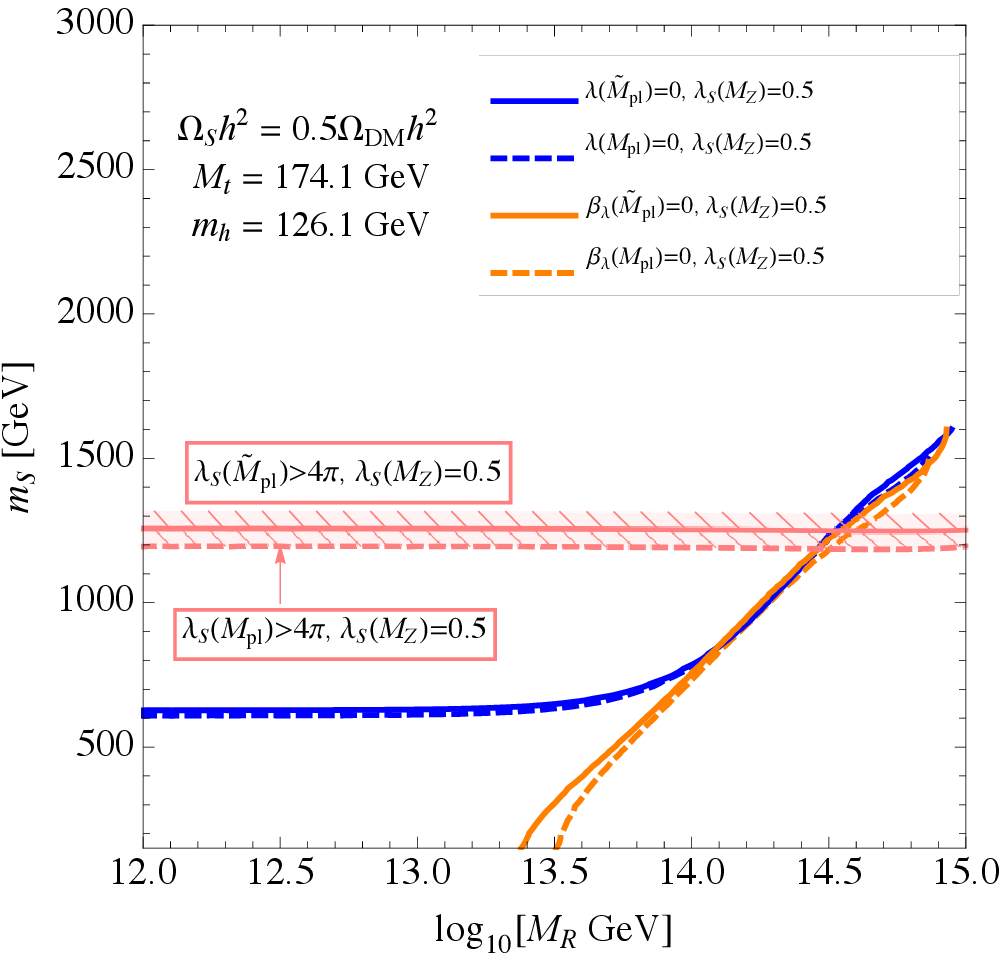} \\
(e) & (f) \\
\includegraphics[scale=0.45]{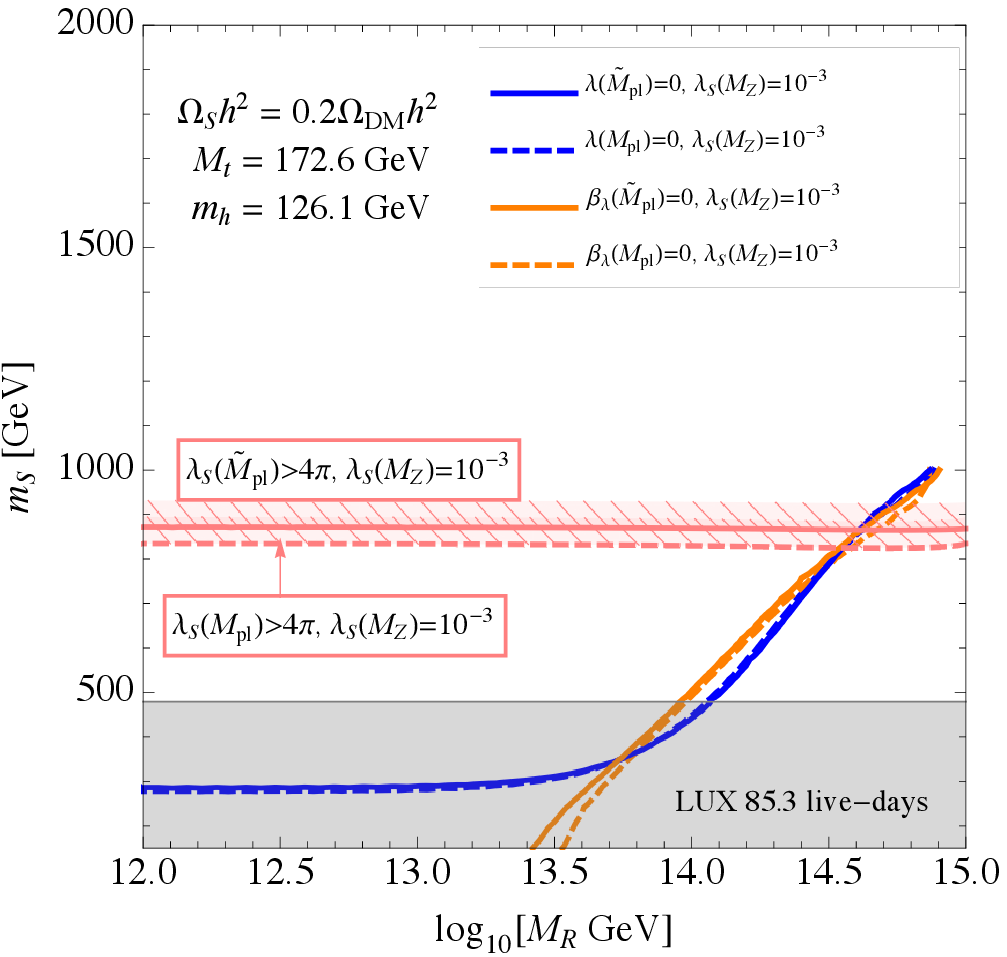} &
\includegraphics[scale=0.45]{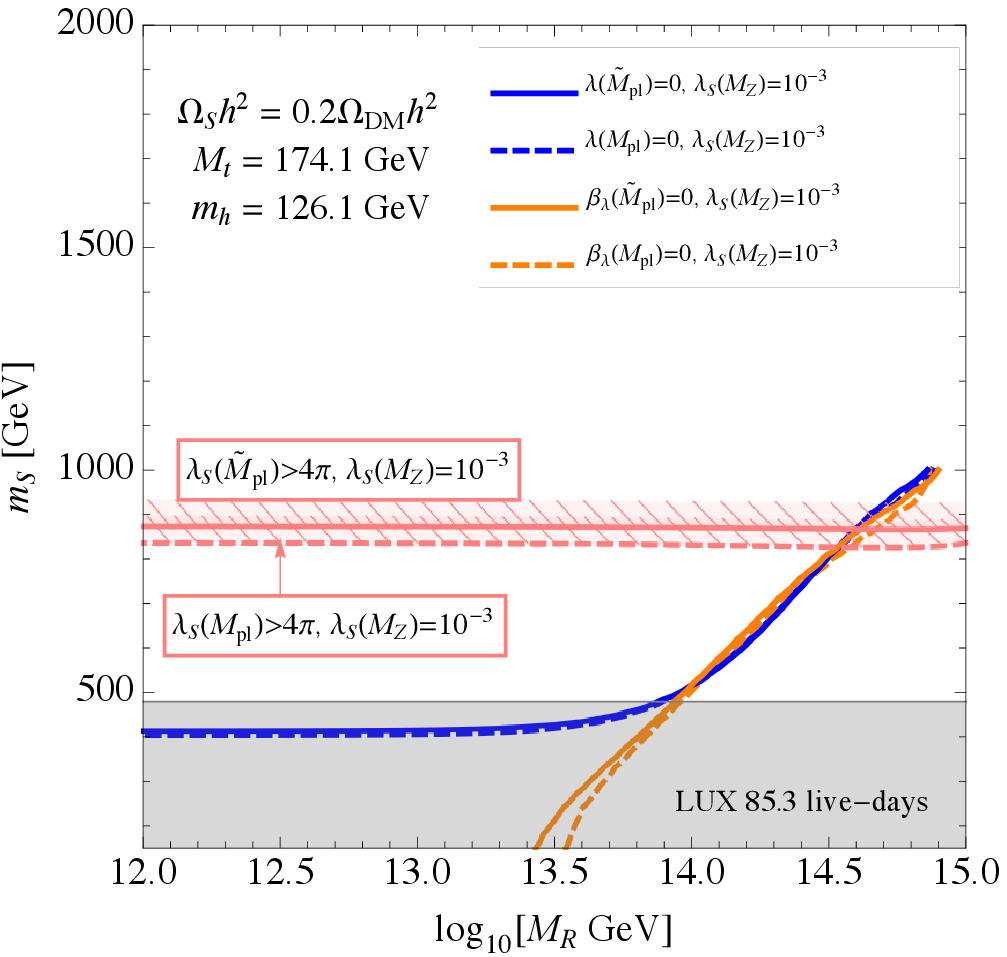} \\
(g) & (h) \\
\includegraphics[scale=0.45]{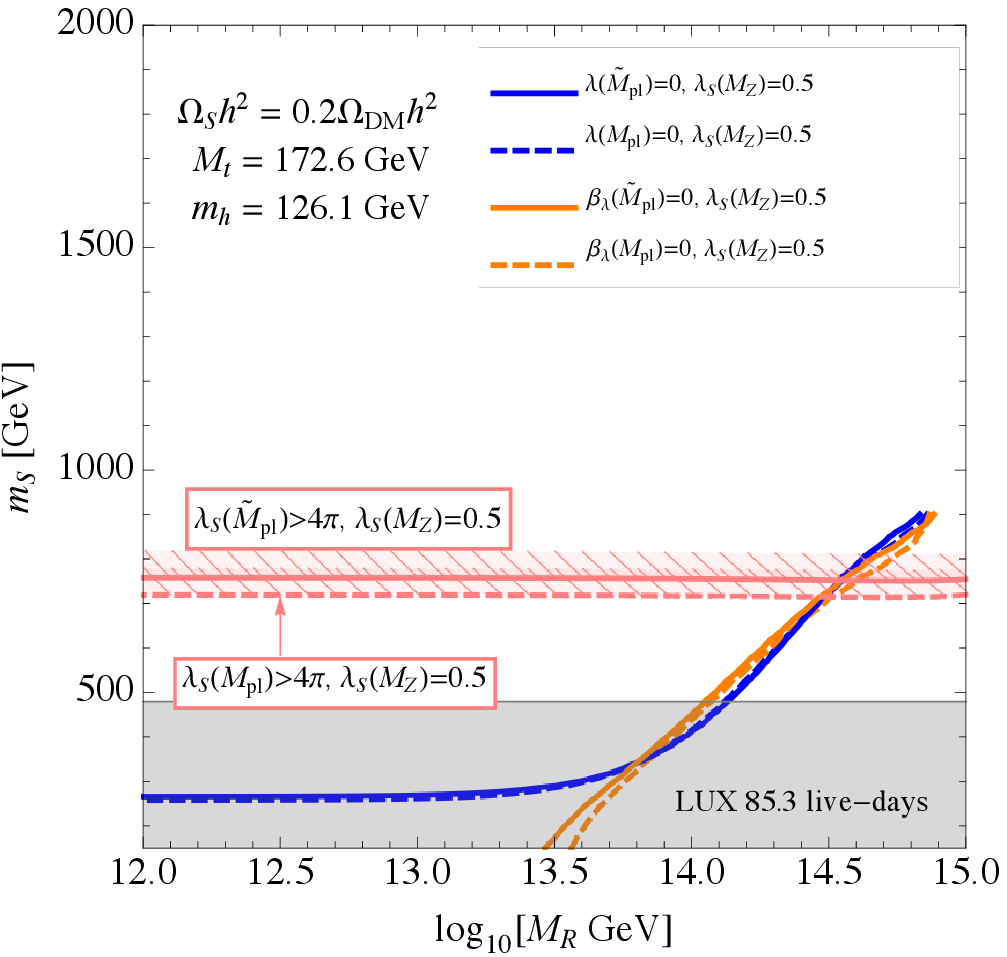} &
\includegraphics[scale=0.45]{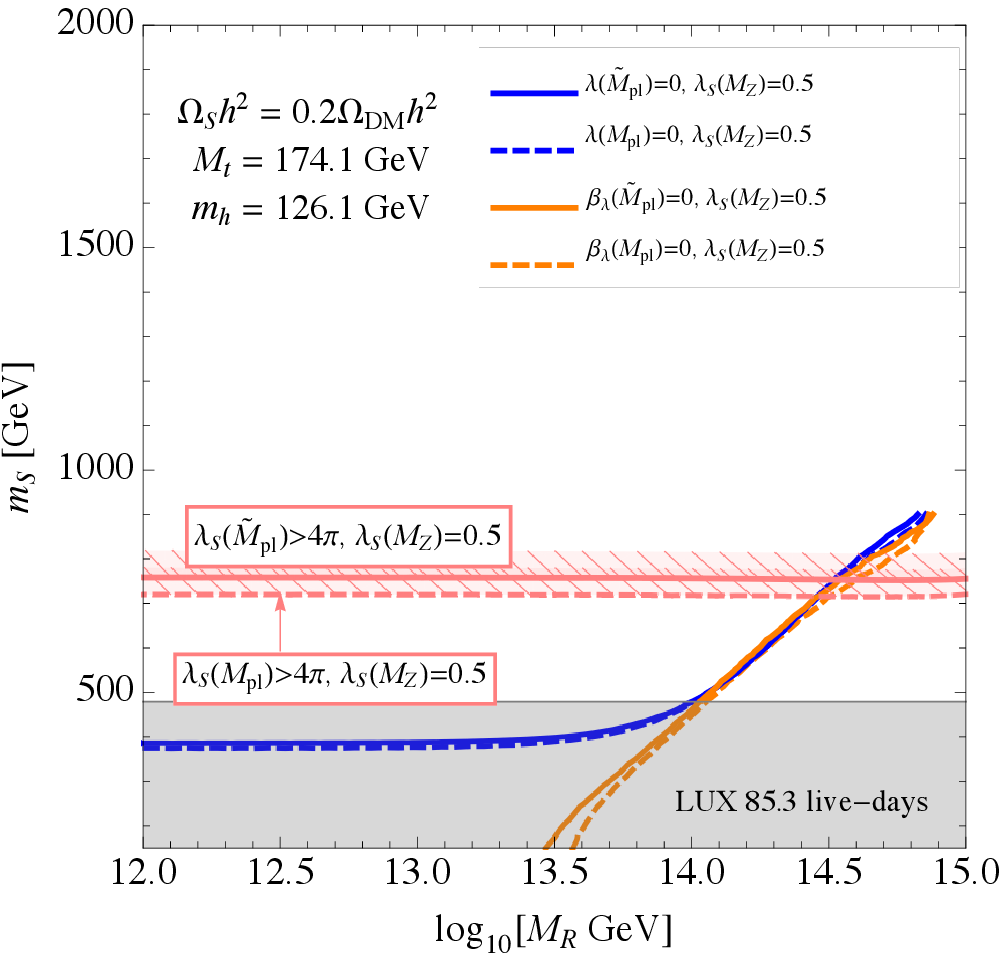} 
\end{tabular}
\end{center}
\caption{Numerical results for the realization of the MPCP [$\lambda=0$ (blue 
curve) and $\beta_\lambda=0$ (orange curve)] in the cases of $\Omega_S/\Omega_{\rm 
DM}=0.5$ [(a)-(d)] and $\Omega_S/\Omega_{\rm DM}=0.2$ [(e)-(h)]. The meanings of 
the figures are the same as in Fig.~\ref{fig1}.}
\label{fig1-2}
\end{figure}

Finally, we also show other solutions of the MPCP as examples of different shares of $\Omega_S$ for $\Omega_{\rm DM}$; i.e., $\Omega_S/\Omega_{\rm DM}=0.5$, shown in Figs.~\ref{fig1-2}(a)-\ref{fig1-2}(d), and 0.2 in Figs.~\ref{fig1-2}(e)-\ref{fig1-2}(h). 
The meanings of the lines and colors are the same as in Fig.~\ref{fig1}. 
One can see that the MPCP 
can be realized in a lighter $m_S$ region compared to the $\Omega_S/\Omega_{\rm DM}=1$
 case. At the same time, the bound from the coupling perturbativity of 
$\lambda_S$ on $m_S$ becomes more severe when the value of $\Omega_S/\Omega_{\rm 
DM}$ becomes smaller than unity (see Figs.~\ref{fig1} and \ref{fig1-2}). This 
is because a smaller $\Omega_S$ needs a larger value of $k(M_Z)$ for the same DM 
mass [e.g., see the green ($\Omega_S/\Omega_{\rm DM}=0.5$) and the yellow 
($\Omega_S/\Omega_{\rm DM}=0.2$) dashed curves in Fig.~\ref{fig2}]. Thus, a lighter 
$m_S$ gives a solution for the MPCP, and a heavier $m_S$ region is constrained by 
the coupling perturbativity of $\lambda_S$ in a smaller $\Omega_S/\Omega_{\rm DM}$
 case. We also show excluded (shaded) regions of $m_S<480$ GeV by the LUX 85.3 
live-day WIMP search~\cite{Akerib:2013tjd} in Figs.~\ref{fig1-2}(e)-\ref{fig1-2}(h). 
Regarding an experimental bound on DM, although there are intersection 
points around $m_S\sim400$ GeV in the case of $\Omega_S/\Omega_{\rm DM}=0.2$ with 
$M_t=172.6$ GeV [see Figs.~\ref{fig1-2}(e) and \ref{fig1-2}(g)], the LUX experiment has 
ruled out $m_S<480$ GeV. As a result, the MPCP is satisfied in the regions 
$5.5\times10^2~{\rm GeV}\leq m_S\leq8.4\times10^2~{\rm GeV}$ and 
$6.3\times10^{13}~{\rm GeV}\leq M_R\leq1.6\times10^{14}~{\rm GeV}$ for 
$\Omega_S/\Omega_{\rm DM}=0.5$ within $M_t=(172.6\mathchar`-174.1)$ GeV, and 
$m_S=5.1\times10^2~{\rm GeV}$ and $M_R=1.0\times10^{14}~{\rm GeV}$ for 
$\Omega_S/\Omega_{\rm DM}=0.2$ with $M_t=174.1$ GeV. One can find that these $m_S$
 regions for the realization of the MPCP can also be consistent with the current
 DM direct detection experiments, and they will be checked by future DM direct 
searches. 

\subsection*{Acknowledgement}

This work is partially supported by Scientific Grant by the Ministry of Education 
and Science, No. 24540272. The work of R.T. is supported by research 
fellowships of the Japan Society for the Promotion of Science for Young 
Scientists.

\end{document}